%% file: main.tex
\documentclass[runningheads,12pt]{llncs}

\input{common.tex}

\begin{document}

\title{A parameterized approximation algorithm
  for the Multiple Allocation $k$-Hub Center%
  \thanks{Research supported by São Paulo
  Research Foundation (FAPESP), grants
    \#2015/11937-9 and  
    \#2019/10400-2      
  and National Council for Scientific and
  Technological Development (CNPq), grants
    \#422829/2018-8 and 
    \#312186/2020-7.}   
}

\author{
  Marcelo P. L. Benedito 
  \and
  Lucas P. Melo 
  \and
  Lehilton L. C. Pedrosa 
}
\authorrunning{Benedito et al.}

\institute{
  Institute of Computing, University of Campinas, Brazil \\
  \email{\{mplb,lehilton\}@ic.unicamp.br},
  \email{lucaspradomelo@gmail.com}
}

\maketitle

\input{content/abstract.tex}
\input{content/introduction.tex}
\input{content/preliminaries.tex}
\input{content/hardness.tex}
\input{content/algorithm.tex}
\input{content/analysis.tex}
\input{content/remarks.tex}

\bibliographystyle{splncs04}
\bibliography{main.bib}

\clearpage

\appendix
\input{content/planar.tex}

\end{document}

%% file: common.tex
\usepackage[english]{babel}
\usepackage{xspace}
\usepackage{amsfonts}
\usepackage{amsmath}
\usepackage{mathtools}
\usepackage[hidelinks]{hyperref}
\usepackage[inline,shortlabels]{enumitem}
\usepackage{tikz}

\usetikzlibrary{shapes}
\usetikzlibrary{positioning}
\usetikzlibrary{calc}
\usetikzlibrary{fit}
\usetikzlibrary{chains}

\tikzset{hub/.style={rectangle, fill=darkgray!20, draw, minimum size=0.8cm, inner sep=0pt}}
\tikzset{client/.style={circle, fill=white!10, draw, minimum size=0.8cm, inner sep=0pt}}
\tikzset{vertex/.style={circle, fill=white!15, draw, minimum size=0.5cm, inner sep=0pt}}
\tikzset{smallcircle/.style={circle, draw=black, fill=white, inner sep=0pt, minimum size=5pt}}
\tikzset{smallrect/.style={rectangle, draw=black, fill=white, inner sep=0pt, minimum size=5pt}}
\tikzset{smalldiam/.style={diamond, draw=black, fill=white, inner sep=0pt, minimum size=6pt}}
\tikzset{vertexblack/.style={circle, draw=black, fill=black, inner sep=0pt, minimum size=5pt}}
\tikzset{vertexwhite/.style={circle, draw=black, fill=white, inner sep=0pt, minimum size=5pt}}
\tikzset{vertexgray/.style={circle, draw=black, fill=black!20!white, inner sep=0pt, minimum size=5pt}}
\tikzset{vertexinvisible/.style={circle, draw=white, fill=white, inner sep=0pt, minimum size=5pt}}
\tikzset{hubwhite/.style={rectangle, draw=black, fill=white, inner sep=0pt, minimum size=5pt}}
\tikzset{rectat/.style={rectangle, rounded corners=0.1mm, fill=white, draw, minimum size=0.5cm, inner sep=2pt}}

\newcommand{\integersnn}{\mathbb{Z}_{\geq 0}}

\newcommand{\etal}{\emph{et~al.}\@\xspace}

\newcommand{\OA}{\mathcal{O}^*}
\newcommand{\OO}{\mathcal{O}}

\newcommand{\p}{\textup{P}\xspace}
\newcommand{\np}{\textup{NP}\xspace}

\newcommand{\fpt}{\textup{FPT}\xspace}
\newcommand{\wone}{\textup{W[1]}\xspace}
\newcommand{\wtwo}{\textup{W[2]}\xspace}

\newcommand{\opt}{\textup{OPT}\xspace}

\newcommand{\tw}{\textup{tw}\xspace}
\newcommand{\pw}{\textup{pw}\xspace}

\newcommand{\vcfull}{\textsc{Vertex Cover}\xspace}
\newcommand{\kcfull}{\textsc{$k$-Center}\xspace}
\newcommand{\ksfull}{\textsc{$k$-Supplier}\xspace}
\newcommand{\hsfull}{\textsc{Hitting Set}\xspace}
\newcommand{\makhcfull}{\textsc{Multiple Allocation $k$-Hub Center}\xspace}

\newcommand{\vcc}{\textup{VC}\xspace}
\newcommand{\hs}{\textup{HS}\xspace}
\newcommand{\makhc}{\textup{MA$k$HC}\xspace}

\newcommand{\universe}{\mathcal{U}}
\newcommand{\family}{\mathcal{F}}

\newcommand{\clients}{\mathcal{C}}
\newcommand{\demands}{\mathcal{D}}
\newcommand{\hubs}{\mathcal{H}}

\newcommand{\tree}{\mathcal{T}}
\newcommand{\troot}{{t_{\scriptscriptstyle 0}}}

\newcommand{\cempty}{{c_{\scriptscriptstyle \emptyset}}}
\newcommand{\ca}{\bar{c}}
\newcommand{\ce}{c^*}
\newcommand{\De}{D^*}

\newcommand{\upa}[1]{\uparrow\hspace{-3pt}#1}
\newcommand{\downa}[1]{\downarrow\hspace{-3pt}#1}

\newcommand{\cone}{\textup{\sffamily\small(C1)}\xspace}
\newcommand{\ctwo}{\textup{\sffamily\small(C2)}\xspace}

\usepackage{geometry}
\usepackage{setspace}


%% file: content/abstract.tex
\begin{abstract}
  In the \makhcfull (\makhc), we are given a connected edge-weighted graph~$G$,
  sets of clients~$\clients$ and hub locations~$\hubs$, where ${V(G) = \clients
      \cup \hubs}$, a set of demands ${\demands \subseteq \clients^2}$ and a
  positive integer $k$.
  A solution is a set of hubs ${H \subseteq \hubs}$ of size $k$ such that every
  demand $(a,b)$ is satisfied by a path starting in~$a$, going through some
  vertex of $H$, and ending in~$b$.
  The objective is to minimize the largest length of a path.
  We show that finding a $(3-\epsilon)$-approximation is \np-hard already for
  planar graphs. For arbitrary graphs, the approximation lower bound holds even
  if we parameterize by $k$ and the value $r$ of an optimal solution.
  An exact \fpt~algorithm is also unlikely when the parameter combines $k$ and
  various graph widths, including pathwidth.
  To confront these hardness barriers, we give a $(2+\epsilon)$-approximation
  algorithm parameterized by treewidth, and, as a byproduct, for unweighted
  planar graphs, we give a $(2+\epsilon)$-approximation algorithm parameterized
  by~$k$ and $r$.
  Compared to classical location problems, computing the length of a path
  depends on non-local decisions.
  This turns standard dynamic programming algorithms impractical, thus our
  algorithm approximates this length using only local information.
  We hope these ideas find application in other problems with similar cost
  structure.
\end{abstract}

\keywords{parameterized approximation algorithm \and hub location problem \and treewidth}

%% file: content/introduction.tex
\section{Introduction}
\label{section:introduction}

In the classical location theory, the goal is to select a set of centers or
facilities to serve a set of
clients~\cite{Hochbaum82b,CharikarGTS02,HochbaumS85,CornuejolsNW83}. Usually,
each client is simply connected to the closest selected facility, so that the
transportation or connection cost is minimized.
In several scenarios, however, the demands correspond to connecting a set of
pair of clients.
Rather than connecting each pair directly, one might select a set of hubs that
act as consolidation points to take advantage of economies of
scale~\cite{KaraT00,CampbellEK02,GelarehP11,KarimiB11}.
In this case, each origin-destination demand is served by a path starting at the
origin, going through one or more selected hubs and ending at the destination.
Using consolidation points reduces the cost of maintaining the network, as a
large number of goods is often transported through few hubs, and a small fleet
of vehicles is sufficient to serve the network~\cite{Campbell96}.

Many hub location problems have emerged through the years, that vary depending
on the solution domain, whether it is discrete or continuous; on the number of
hub stops serving each demand; on the number of selected hubs, and so
on~\cite{AlumurK08,FarahaniHAN13}.
Central to this classification is the nature of the objective function: for
\emph{median} problems, the objective is to minimize the total length of the
paths serving the demands, while, for \emph{center} problems, the objective is
to find a solution whose maximum length is minimum.
In this paper, we consider the \makhcfull (\makhc), which is a center problem
in the one-stop model~\cite{JailletSY96,Yang09}, where clients may be assigned
to multiple hubs for distinct demands, and whose objective is to select $k$ hubs
to minimize the worst connection cost of a demand.

Formally, an instance of \makhc is comprised of a connected edge-weighted graph
$G$, sets of clients~$\clients$ and hub locations~$\hubs$, where $V(G) =
\clients \cup \hubs$, a set of demand pairs $\demands \subseteq \clients^2$ and
a positive integer $k$.
The objective is to find a set of hubs $H \subseteq \hubs$ of size $k$ that
minimizes ${\max_{(a,b) \in \demands} \min_{h \in H} d(a,h) + d(h,b)}$,
where $d(u,v)$ denotes the length of a shortest path between vertices
$u$~and~$v$.
In the decision version of \makhc, we are also given a non-negative number~$r$,
and the goal is to determine whether there exists a solution of value at
most~$r$.

This problem is closely related to the well-known
\kcfull~\cite{HochbaumS85,Gonzalez85}, where, given an edge-weighted graph $G$,
one wants to select a set of $k$ vertices, called centers, so that the maximum
distance from each vertex to the closest center is minimized. In the
corresponding decision version, one also receives a number $r$, and asks whether
there is a solution of value at most $r$.
By creating a demand $(u,u)$ for each vertex $u$ of $G$, one reduces \kcfull to
\makhc, thus \makhc can be seen as a generalization of \kcfull. In fact, \makhc
even generalizes the \ksfull~\cite{HochbaumS86}, that is a variant of \kcfull
whose vertices are partitioned into clients and locations, only clients need to
be served, and centers must be selected from the set of locations.

For \np-hard problems, one might look for an $\alpha$-approximation, that is a
polynomial-time algorithm that finds a solution whose value is within a factor
$\alpha$ of the optimal.
For \kcfull, a simple greedy algorithm already gives a $2$-approximation, that
is the best one can hope for, since finding an approximation with smaller factor
is $\np$-hard~\cite{Gonzalez85}.
Analogously, there is a best-possible $3$-approximation for
\ksfull~\cite{HochbaumS86}.
These results have been extended to \makhc as well, which also admits a
$3$-approximation~\cite{PedrosaSS16}. Later, we prove this approximation factor
is tight, unless ${\p = \np}$.

An alternative is to consider the problem from the perspective of parameterized
algorithms, that insist on finding an exact solution, but allow running times
with a non-polynomial factor that depends only on a certain parameter of the
input. More precisely, a decision problem with parameter $w$ is
\emph{fixed-parameter tractable} (\fpt) if it can be decided in time $f(w) \cdot
n^{O(1)}$, where $n$ is the size of the input and $f$ is a function that depends
only on $w$.
Feldmann and Marx~\cite{FeldmannM20} showed that \kcfull is \wone-hard for
planar graphs of constant doubling dimension when the parameter is a combination
of $k$, the highway dimension and the pathwidth of the graph. Blum~\cite{Blum21}
showed that the hardness holds even if we additionally parameterize by the
skeleton dimension of the graph.
Under the assumption that $\fpt \ne \wone$, this implies that \kcfull does not
admit an \fpt~algorithm for any of these parameters, even if restricted to
planar graphs of constant doubling dimension.

Recently, there has been interest in combining techniques from parameterized and
approximation algorithms~\cite{Marx08,FeldmannKLM20}.
An algorithm is called a parameterized $\alpha$-approximation if it finds a
solution within factor $\alpha$ of the optimal value and runs in \fpt~time.
The goal is to give an algorithm with improved approximation factor that runs in
super-polynomial time, where the non-polynomial factors of the running time are
dependent on the parameter only.
Thus, one may possibly design an algorithm that runs in \fpt~time for a
\wone-hard problem that, although it finds only an approximate solution, has an
approximation factor that breaks the known \np-hardness lower bounds.

For \kcfull, Demaine~\etal~\cite{DemaineFHT05} give an \fpt~algorithm
parameterized by $k$ and $r$ for planar and map graphs. All these
characteristics seem necessary for an exact \fpt~algorithm, as even finding a
$(2-\epsilon)$-approximation with $\epsilon > 0$ for the general case is
\wtwo-hard for parameter $k$~\cite{Feldmann19}.
If we remove the solution value~$r$ and parameterize only by $k$, the problem
remains \wone-hard if we restrict the instances to planar
graphs~\cite{FeldmannM20}, or if we add structural graph parameters, such as the
vertex-cover number or the feedback-vertex-set number (and thus, also treewidth
or pathwidth)~\cite{KatsikarelisLP19}.

To circumvent the previous barriers, Katsikarelis~\etal~\cite{KatsikarelisLP19}
provide an efficient parameterized approximation scheme (EPAS) for \kcfull with
different parameters~$w$, i.e., for every $\epsilon > 0$, one can compute a $(1
+ \epsilon)$-approximation in time $f(\epsilon, w) \cdot n^{O(1)}$, where $w$ is
either the cliquewidth or treewidth of the graph.
More recently, Feldmann and Marx~\cite{FeldmannM20} have also given an EPAS for
\kcfull when it is parameterized by $k$ and the doubling dimension, which can be
a more appropriate parameter for transportation networks than~$r$.

\paragraph{Our results and techniques}

We initiate the study of \makhc under the perspective of parameterized
algorithms.
We start by showing that, for any $\epsilon > 0$, there is no parameterized
$(3-\epsilon)$-approximation for \makhc when the parameter is $k$, the value $r$
is bounded by a constant and the graph is unweighted, unless $\fpt = \wtwo$.
For planar graphs, finding a good constant-factor approximation remains hard in
the polynomial sense, as we show that it is \np-hard to find a
$(3-\epsilon)$-approximation for \makhc in this case, even if the maximum degree
is 3.

To challenge the approximation lower bound, one might envisage an \fpt
algorithm by considering an additional structural parameter, such as
vertex-cover and feedback-vertex-set numbers or treewidth.
However, this is unlikely to lead to an exact \fpt~algorithm, as we note that
the hardness results for \kcfull~\cite{KatsikarelisLP19,FeldmannM20,Blum21}
extend to \makhc. Namely, we show that, unless $\fpt = \wone$, \makhc does not
admit an \fpt~algorithm when parameterized by a combination of $k$, the highway
and skeleton dimensions and the pathwidth of the graph, even if restricted to
planar graphs of constant doubling dimension; or when parameterized by $k$ and
the vertex-cover number.
Instead, we aim at finding an approximation with factor strictly smaller
than~$3$ that runs in \fpt~time.

In this paper, we present a $(2+\epsilon)$-approximation for \makhc
parameterized by the treewidth of the graph, for $\epsilon > 0$. The running
time of the algorithm is $\OA((\tw/\epsilon)^{\OO(\tw)})$, where polynomial
factors in the size of the input are omitted.
Moreover, we give a parameterized ${(2+\epsilon)}$-approximation for \makhc when
the input graph is planar and unweighted, parameterized by~$k$ and $r$.

Our main result is a non-trivial dynamic programming algorithm over a tree
decomposition, that follows the spirit of the algorithm by
Demaine~\etal~\cite{DemaineFHT05}.
We assume that we are given a tree decomposition of the graph and consider both
$k$ and $r$ as part of the input. Thus, for each node $t$ of this decomposition,
we can guess the distance from each vertex in the bag of $t$ to its closest hub
in some (global) optimal solution~$H^*$.
The subproblem is computing the minimum number of hubs to satisfy each demand in
the subgraph $G_t$, corresponding to $t$.

Compared to \kcfull and \ksfull, however, \makhc has two additional sources of
difficulty.
First, the cost to satisfy a demand cannot be computed locally, as it is the sum
of two shortest paths, each from a client in the origin-destination pair to some
hub in $H^*$ that satisfies that pair. Second, the set of demand pairs
$\demands$ is given as part of the input, whereas every client must be served in
\kcfull or in \ksfull.
If we knew the subset of demands $\De_t$ that are satisfied by some hub in $H^*
\cap V(G_t)$, then one could solve every subproblem in a bottom-up fashion, so
that every demand would have been satisfied in the subproblem corresponding to
the root of the decomposition.

Guessing $\De_t$ leads to an \fpt~algorithm parameterized by $\tw$, $r$ and
$|\demands|$, which is unsatisfactory as the number of demands might be large in
practice.
Rather, for each node $t$ of the tree decomposition, we compute
deterministically two sets of demands $D_t,S_t \subseteq \demands$ that enclose
$\De_t$, that is, that satisfy ${D_t \subseteq \De_t \subseteq D_t \cup S_t}$.
By filling the dynamic programming table using $D_t$ instead of $\De_t$, we can
obtain an algorithm that runs in \fpt~time on parameters $\tw$ and $r$, and that
finds a $2$-approximation.

The key insight for the analysis is that the minimum number of hubs in $G_t$
that are necessary to satisfy each demand in $D_t$ by a path of length at most
$r$ is a lower bound on $|H^* \cap V(G_t)|$. At the same time, the definition of
the set of demands~$S_t$ ensures that each such demand can be satisfied by a
path of length at most $2r$ using a hub that is close to a vertex in the bag of
$t$.
This is the main technical contribution of the paper, and we believe that these
ideas might find usage in algorithms for similar problems whose solution costs
have non-local components.

Using only these ideas, however, is not enough to get rid of $r$ as a parameter,
as we need to enumerate the distance from each vertex in a bag to its closest
hub.
A common method to shrink a dynamic programming table with large integers is
storing only an approximation of each number, causing the solution value to be
computed approximately. This eliminates the parameter $r$ from the running time,
but adds a term $\epsilon$ to the approximation factor.
This technique is now standard~\cite{Lampis14} and has been applied multiple
times for graph width
problems~\cite{DemaineFHT05,FominGLS10b,KatsikarelisLP19,BeneditoP21}.

Specifically, we employ the framework of approximate addition
trees~\cite{Lampis14}.
For some ${\delta > 0}$, we approximate each value $\{1,\dots,r\}$ of an entry
in the dynamic programming table by an integer power of $(1+\delta)$, and show
that each such value is computed by an addition tree and corresponds to an
approximate addition tree.
By results in~\cite{Lampis14}, we can readily set $\delta$ appropriately so that
the number of distinct entries is polynomially bounded and each value is
approximated within factor $(1+\epsilon)$.

\paragraph{Related work}
The first modern studies on hub location problems date several decades back,
when models and applications were surveyed~\cite{OKelly86,OKelly87}. Since then,
most papers focused on integer linear programming and heuristic
methods~\cite{AlumurK08,FarahaniHAN13}.
Approximation algorithms were studied for the single allocation median variant,
whose task is to allocate each client to exactly one of the given hubs,
minimizing the total transportation cost~\cite{IwasaSM09,AndoM11,GeHYZ10}.
Later, constant-factor approximation algorithms were given for the problem of,
simultaneously, selecting hubs and allocating clients~\cite{BeneditoP19}.
The analogous of \makhc with median objective was considered by Bordini and
Vignatti~\cite{BordiniV17}, who presented a $(4\alpha)$-approximation algorithm
that opens $\Big(\frac{2\alpha}{2\alpha - 1}\Big)k$ hubs, for $\alpha > 1$.

There is a single allocation center variant that asks for a two-level hub
network, where every client is connected to a single hub and the path satisfying
a demand must cross a given network center~\cite{YamanE12,Liang13}.
Chen~\etal~\cite{ChenCHHLW16} give a $\frac{5}{3}$-approximation algorithm and
showed that finding a $(1.5-\epsilon)$-approximation, for $\epsilon > 0$, is
\np-hard.
This problem was shown to admit an EPAS parameterized by the
treewidth~\cite{BeneditoP21} and, to our knowledge, is the first hub location
problem studied in the parameterized setting.

\paragraph{Organization}

The remainder of the paper is organized as follows.
Section~\ref{section:preliminaries} introduces basic concepts and describes the
framework of approximate addition trees.
Section~\ref{section:hardness} shows the hardness results for \makhc in both
classical and parameterized complexity.
Section~\ref{section:algorithm} presents the approximation algorithm
parameterized by treewidth, which is analyzed in
Section~\ref{section:analysis}.
Section~\ref{section:remarks} presents the final remarks.
The case of planar graphs is considered in Appendix~\ref{section:planar}.

%% file: content/preliminaries.tex
\section{Preliminaries}
\label{section:preliminaries}

An \emph{$\alpha$-approximation algorithm} for a minimization problem is an
algorithm that, for every instance $I$ of size $n$, has running time
$n^{\OO(1)}$ and outputs a solution of value at most $\alpha \cdot \opt(I)$,
where $\opt(I)$ is the optimal value of $I$.
A \emph{parameterized algorithm} for a parameterized problem is an algorithm
that, for every instance~$(I, k)$, has running time $f(k) \cdot n^{\OO(1)}$,
where $f$ is a computable function that depends only on the parameter~$k$, and
decides $(I,k)$ correctly.
A parameterized problem that admits a parameterized algorithm is called
\emph{fixed-parameter tractable}, and the set of all such problems is denoted
by~\fpt.
Finally, a \emph{parameterized $\alpha$-approximation algorithm} for a
(parameterized) minimization problem is an algorithm that, for every instance
$I$ and corresponding parameter $k$, has running time ${f(k) \cdot n^{\OO(1)}}$
and outputs a solution of value at most $\alpha \cdot \opt(I)$.
For a complete exposition, we refer the reader
to~\cite{WilliamsonS11,CyganFKLMPPS15,Marx08}.

We adopt standard graph theoretic notation.
Given a graph $G$, we denote the set of vertices and edges as $V(G)$ and $E(G)$,
respectively. For $S \subseteq V(G)$, the subgraph of $G$ induced by $S$ is
denoted as $G[S]$ and is composed by the vertices of $S$ and every edge of the
graph that has both endpoints in $S$.

A \emph{tree decomposition} of a graph $G$ is a pair $(\tree,X)$, where $\tree$
is a tree and $X$ is a function that associates a node $t$ of $\tree$ to a set
$X_t \subseteq V(G)$, called \emph{bag}, such that:
\begin{enumerate}[$(i)$]

  \item $\cup_{t \in V(\tree)}X_t = V(G)$;

  \item for every $(u,v) \in E(G)$, there exists $t \in V(\tree)$ such that
    $u,v \in X_t$;

  \item for every $u \in V(G)$, the set $\{t \in V(\tree): u \in X_t\}$ induces
    a connected subtree of $\tree$.

\end{enumerate}

The width of a tree decomposition is $\max_{t \in V(\tree)}|X_t|-1$ and the
\emph{treewidth} of~$G$ is the minimum width of any tree decomposition of the
graph.
Also, for a node $t \in V(\tree)$, let $\tree_t$ be the subset of nodes that
contains $t$ and all its descendants, and define $G_t$ as the induced subgraph
of $G$ that has $\bigcup_{t' \in \tree_t} X_{t'}$ as the set of vertices.

Dynamic programming algorithms over tree decompositions often assume that the
decomposition has a restricted structure.
In a \emph{nice tree decomposition} of $G$, $\tree$ is a binary tree and each
node $t$ has one of the following types:
\begin{enumerate}[$(i)$]

  \item \emph{leaf node}, which has no child and $X_t = \emptyset$;

  \item \emph{introduce node}, which has a child $t'$ with $X_t = X_{t'}\cup
    \{u\}$, for $u \notin X_{t'}$;

  \item \emph{forget node}, which has a child $t'$ with $X_t =
    X_{t'}\setminus \{u\}$, for $u \in X_{t'}$;

  \item \emph{join node}, which has children $t'$ and $t''$ with $X_t =
    X_{t'} = X_{t''}$.

\end{enumerate}

Given a tree decomposition $(\tree,X)$ of width $\tw$, there is a
polynomial-time algorithm that finds a nice tree decomposition of the same width
and ${\OO(\tw \cdot |V(G)|)}$ nodes~\cite{Kloks94}.
Moreover, we may assume without loss of generality that our algorithm receives
as input a nice tree decomposition of $G$ whose tree has height $\OO(\tw \cdot
\log |V(G)|)$, using the same arguments as discussed
in~\cite{BodlaenderH98,BeneditoP21}.

\subsection{Approximate addition trees}

An \emph{addition tree} is an abstract model that represents the computation of
a number by successively adding two other previously computed numbers.

\begin{definition}
  An \emph{addition tree} is a full binary tree such that each leaf $u$ is
  associated to a non-negative integer input $y_u$, and each internal node $u$
  with children $u'$ and $u''$ is associated to a computed number $y_u \coloneqq
  y_{u'} + y_{u''}$.
\end{definition}

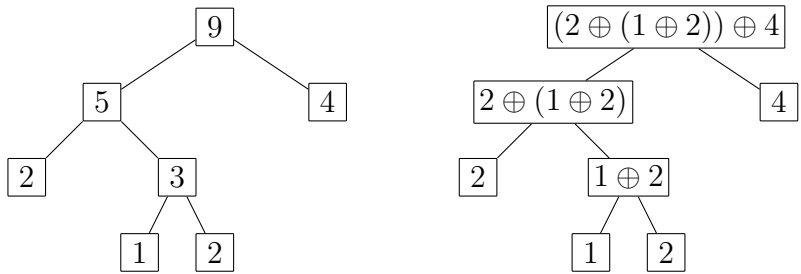
\begin{figure}
  \centering
  \begin{tikzpicture}
    \node[rectat] at (0, 0) (at0) {$9$};
    \node[rectat] at (-1.5, -1) (at1) {$5$};
    \node[rectat] at (1.5, -1) (at2) {$4$};
    \node[rectat] at (-2.5, -2) (at3) {$2$};
    \node[rectat] at (-0.5, -2) (at4) {$3$};
    \node[rectat] at (-1, -3) (at5) {$1$};
    \node[rectat] at (0, -3) (at6) {$2$};

    \node[rectat] at (6 + 0, 0) (aat0) {$(2 \oplus (1 \oplus 2)) \oplus 4$};
    \node[rectat] at (6 + -1.5, -1) (aat1) {$2 \oplus (1 \oplus 2)$};
    \node[rectat] at (6 + 1.5, -1) (aat2) {$4$};
    \node[rectat] at (6 + -2.5, -2) (aat3) {$2$};
    \node[rectat] at (6 + -0.5, -2) (aat4) {$1 \oplus 2$};
    \node[rectat] at (6 + -1, -3) (aat5) {$1$};
    \node[rectat] at (6 + 0, -3) (aat6) {$2$};

    \path[draw]
    (at0) edge node{} (at1)
    (at0) edge node{} (at2)
    (at1) edge node{} (at3)
    (at1) edge node{} (at4)
    (at4) edge node{} (at5)
    (at4) edge node{} (at6)
    (aat0) edge node{} (aat1)
    (aat0) edge node{} (aat2)
    (aat1) edge node{} (aat3)
    (aat1) edge node{} (aat4)
    (aat4) edge node{} (aat5)
    (aat4) edge node{} (aat6);

  \end{tikzpicture}
  \caption{Example calculation in both types of Trees.}
  \label{figure:addition_trees}
\end{figure}

One can replace the sum with some operator $\oplus$, which computes each such
sum only approximately, up to an integer power of $(1+\delta)$, for some
parameter ${\delta > 0}$. The resulting will be an \emph{approximate addition
tree}.
While the error of the approximate value can pile up as more operations are
performed, Lampis~\cite{Lampis14} showed that, for some $\epsilon > 0$, as long
as $\delta$ is not too large, the relative error can bounded by $1 + \epsilon$.
Figure~\ref{figure:addition_trees} illustrates an addition tree and the
corresponding approximate addition tree.

\begin{definition}
  An \emph{approximate addition tree} with parameter $\delta > 0$ is a full
  binary tree, where each leaf $u$ is associated to a non-negative integer input
  $z_u$, and each internal node $u$ with children $u'$ and $u''$ is associated
  to a computed value ${z_u \coloneqq z_{u'} \oplus z_{u''}}$, where $a \oplus b
  \coloneqq 0$ if both $a$ and $b$ are zero, and $a \oplus b \coloneqq
  (1+\delta)^{\lceil\log_{1+\delta}(a+b)\rceil}$, otherwise.
\end{definition}

For simplicity, here we defined only a deterministic version of the approximate
addition tree, since we can assume that the height of the tree decomposition is
bounded by $\OO(\tw \cdot \log |V(G)|)$.
For this case, Lampis showed the following result.

\begin{theorem}[\cite{Lampis14}]
  \label{theorem:lampis}
  Given an approximate addition tree of height~$\ell$, if ${\delta <
  \frac{\epsilon}{2\ell}}$, then, for every node $u$ of the tree, we have
  $\max\Big\{\frac{z_u}{y_u},\frac{y_u}{z_u}\Big\} < 1 + \epsilon$.
\end{theorem}

\subsection{Preprocessing}

For an instance of \makhc and a demand $(a,b) \in \demands$, define $G_{ab}$ as
the induced subgraph of $G$ with vertex set $V(G_{ab}) = \{v \in V(G): d(a, v) +
d(v, b) \le r\}$.

Notice that if a solution $H$ has a hub $h \in V(G_{ab})$, then the length of a
path serving $(a,b)$ that crosses $h$ is at most $r$. In this case, we say that
demand $(a,b)$ is \emph{satisfied} by~$h$ with cost $r$.
Thus, in an optimal solution $H^*$ of \makhc, for every $(a,b) \in \demands$,
the set ${H^* \cap V(G_{ab})}$ must be non-empty.

Also, if there is $v\in V(G)$ such that $d(a,v) + d(v,b) > r$ for every ${(a, b)
    \in \demands}$, then $v$ does not belong to any $(a,b)$-path of length at most
$r$, and can be safely removed from $G$.
From now on, assume that we have preprocessed $G$ in polynomial time, such that
for every~$v \in V(G)$,
\begin{align*}
  \min_{(a,b) \in \demands} d(a,v) + d(v,b) \le r.
\end{align*}

Moreover, we assume that each edge has an integer weight and that the optimal
value, $\opt$, is bounded by $\OO(\frac{1}{\epsilon}|V(G)|)$, for a given
constant $\epsilon > 0$. If not, then we solve another instance for which this
holds and that has optimal value $\opt' \le (1+ \epsilon) \opt$ using standard
rounding techniques~\cite{WilliamsonS11}.
It suffices finding a constant-factor approximation of value $A \le
3\opt$~\cite{PedrosaSS16}, and defining a new distance function such that
$d'(u,v) = \left\lceil \frac{3|V(G)|}{\epsilon A}d(u,v)\right\rceil$.

%% file: content/hardness.tex
\section{Hardness}
\label{section:hardness}

Next, we observe that approximating \makhc is hard, both in the classical and
parameterized senses.
First, we show that approximating the problem by a factor better than $3$ is
\np-hard, even if the input graph is planar and unweighted.
This result strengthens the previous known lower bound and matches the
approximation factor of the greedy algorithm~\cite{PedrosaSS16}.

\begin{theorem}
  \label{theorem:inapprox_in_poly}
  For every $\epsilon > 0$, if there is a $(3-\epsilon)$-approximation for
  \makhc when $G$ is an unweighted planar graph, then $\p = \np$.
\end{theorem}

\begin{proof}
  We present a reduction from \vcfull (\vcc), whose task is to find a subset of
  $k$ vertices that contains at least one endpoint of every edge of the graph.
  More specifically, we consider a particular version of the problem.
  \begin{claim}
    \vcfull is \np-hard even if the input graph is planar, triangle-free and has
    maximum degree 3.
  \end{claim}
  \begin{proof}
    We self-reduce the problem from the case the input graph is planar and with
    maximum degree 3, which is known to be \np-hard~\cite{GareyJ77}.
    Given an instance $(G,k)$ of vertex cover, create another instance
    $(G',k')$, where $G'$ is obtained by subdividing each edge of $G$ in three
    parts, and $k' = k + |E(G)|$.
    Let $u_e$ and $v_e$ be new vertices added for the subdivision of an edge
    $e=(u,v) \in E(G)$ and that are incident with $u$ and $v$, respectively.

    Assume $S$ is a vertex cover for $G$ with size $k$, and build a vertex cover
    $S'$ for $G'$ as follows. Initialize $S'$ with a copy of $S$ and, for each
    edge $e=(u,v)$ of $G$, add $v_e$ to $S'$, if $u \in S$, and add $u_e$,
    otherwise.
    Note that $S'$ is a vertex cover of $G'$ of size~$k'$.
    For the other direction, assume $S'$ is a vertex cover of $G'$ with size~$k'$,
    and define
    $
    S = S' \setminus \{u_e,v_e : e \in E(G)\}.
    $
    If, for some edge $(u_e,v_e)$ of $G'$, both $u_e$ and $v_e$ are in $S'$,
    then $S' \setminus \{u_e\} \cup \{u\}$ is a vertex cover of $G'$.
    Thus, assume for every such edge $(u_e,v_e)$, either $u_e$ or $v_e$ is in
    $S'$. It follows that $S$ is a vertex cover of $G$ of size $k$.\qed
  \end{proof}

  Given an instance $(G,k)$ of \vcc, build an instance $(G, \clients, \hubs,
  \demands, k)$ of \makhc, where~$\clients = \hubs = V(G)$ and $\demands =
  E(G)$. Observe that there exists a vertex cover~$S$ of size~$k$ in~$G$ if, and
  only if, the solution $S$ for \makhc has value 1.
  Suppose that the optimal value is greater than 1, then it would have to be at
  least 3, since the graph has no triangles. Then, for $\epsilon > 0$, a
  $(3-\epsilon)$-approximation for \makhc can decide whether the optimal value
  is 1, thus deciding whether there is a vertex cover of size $k$ in $G$.\qed
\end{proof}

From this reduction, one may observe that the previous theorem holds even for
the case where the maximum degree is~$3$ and the optimal value is bounded
by~$3$.

To find a better approximation guarantee, one might resource to a parameterized
approximation algorithm.
The natural candidates for parameters of \makhc are the number of hubs $k$ and
the value $r$ of an optimal solution.
The next theorem states that this choice of parameters does not help, as it is
\wtwo-hard to find a parameterized approximation with factor better than $3$,
when the parameter is $k$, the value $r$ is bounded by a constant and $G$ is
unweighted.

\begin{theorem}
  \label{theorem:inapprox_in_fpt}
  For every $\epsilon > 0$, if there is a parameterized
  $(3-\epsilon)$-approximation for \makhc with parameter~$k$, then $\fpt =
  \wtwo$.
  This holds even for the particular case of \makhc with instances $I$ such that
  $\opt(I) \le 6$.
\end{theorem}

\begin{proof}
  The theorem will follow by a reduction from \hsfull (\hs), which is known
  to be \wtwo-hard~\cite{DowneyF12}.
  We show that a $(3-\epsilon)$-approximation for \makhc can decide the instance
  of \hs, implying that $\fpt = \wtwo$.
  Remember that in~\hs, we are given a set $\universe$, a family of sets
  $\family \subseteq 2^\universe$ and an integer $k$, and the objective is to
  decide whether there exists a set $H \subseteq \universe$ of size $k$ that
  intersects every set of~$\family$.

  Given an instance $I = (\universe,\family,k)$ of \hs, we build an instance
  $I' = (G,\clients,\hubs,\demands,k)$ of \makhc: for each element $e \in
  \universe$, create a vertex $h_e$ in $G$ and add it to $\hubs$; for each set
  $S \in \family$, create vertices $u_S$ and $v_S$ in $G$, add them to
  $\clients$, create a demand $(u_S,v_S)$ in $\demands$ and connect $u_S$ and
  $v_S$ to vertices $\{h_e : e \in S\}$.

  Consider a hitting set $H$ of size $k$, and let $H' = \{h_e : e \in H\}$ be a
  set of hubs of size $k$.
  This set of hubs satisfies every demand in $\demands$ with cost $2$, since
  for every $S \in \family$, there is $e \in S \cap H$ and thus $h_e \in H'$.
  In the other direction, consider a set of hubs $H'$ of size $k$ that satisfies
  every demand in $\demands$ with cost $2$, and let $H = \{e : h_e \in H'\}$ be
  a set of elements of size~$k$.
  For each set $S \in \family$, there exists a corresponding demand $(u_S,v_S)$
  in $\demands$ that is satisfied by a hub $h_e \in H'$ with cost~$2$. Since the
  length of this path is~$2$, $h_e$ must be a neighbor of $u_S$ and $v_S$
  in~$G$, then $e \in S \cap H$. It follows that $H$ is a hitting set for $I$.

  We have shown that $I$ is a yes-instance if, and only if, the optimal value
  of~$I'$ is~$2$.
  Now, if the optimal value of $I'$ is greater than~$2$, then it would have to
  be at least $6$. Indeed, if a demand $(u_S,v_S)$ is satisfied by a hub $h_e
  \in H'$ with cost greater than~$2$, then $h_e$ is not a neighbor of $u_S$. But
  $G$ is bipartite and $u_S$~and~$h_e$ are at different parts, then ${d(u_S,h_e)
  \geq 3}$. Analogously, we have ${d(v_S,h_e) \geq 3}$, and thus ${d(u_S,h_e) +
  d(v_S,h_e) \geq 6}$.
  We conclude that a $(3-\epsilon)$-approximation can decide whether the optimal
  value of~$I'$~is~$2$, thus deciding whether $I$ is a yes-instance.\qed
\end{proof}

Due to the previous hardness results, a parameterized algorithm for \makhc must
consider different parameters, or assume a particular case of the problem. In
this paper, we focus on the treewidth of the graph, that is one of the most
studied structural parameters~\cite{CyganFKLMPPS15}, and the particular case of
planar graphs. This setting is unlikely to lead to an (exact) \fpt~algorithm,
though, as the problem is \wone-hard, even if we combine these conditions.
The next theorem follows directly from a result of Blum~\cite{Blum21}, since
\makhc is a generalization of \kcfull.

\begin{theorem}
  \label{theorem:w1hardness}
  Even on planar graphs with edge lengths of constant doubling dimension, \makhc
  is \wone-hard for the combined parameter $(k, \pw, h, \kappa)$, where $\pw$ is
  the pathwidth, $h$ is the highway dimension and $\kappa$ is the skeleton
  dimension of the graph.
\end{theorem}

\begin{proof}
  Given an instance $I = (G,k)$ of \kcfull, we build an instance $I' =
  (G,\clients,\hubs,\demands,k)$ of \makhc where $\clients = \hubs = V(G)$ and
  $\demands = \{(u,u) : u \in V(G)\}$.
  Now, note that there is a solution of value at most $r$ for $I$ if, and only
  if, there is a solution of value $2r$ for $I'$. The theorem follows, as we do
  not change the graph or the number of hubs~$k$.\qed
\end{proof}

Note that \makhc inherits other hardness results of \kcfull by
Katsikarelis~\etal~\cite{KatsikarelisLP19}, thus it is \wone-hard when
parameterized by a combination of $k$ and the vertex-cover number.

Recall that the treewidth is a lower bound on the pathwidth, thus the previous
theorem implies that the problem is also \wone-hard for planar graphs when
parameterized by a combination of $k$ and $\tw$.
To circumvent these hardness results, in Section~\ref{section:algorithm}, we
give a $(2 + \epsilon)$-approximation algorithm for \makhc for arbitrary graphs
that is parameterized by $\tw$, breaking the approximation barrier of~$3$.
In Appendix~\ref{section:planar}, we complement this result with a $(2 +
\epsilon)$-approximation for unweighted planar graphs parameterized by~$k$ and
$r$.

%% file: content/algorithm.tex
\section{The algorithm}
\label{section:algorithm}

In this section, we give a $(2+\epsilon)$-approximation parameterized only by
the treewidth.
In what follows, we assume that we receive a preprocessed instance of $\makhc$
and a nice tree decomposition of the input graph $G$ with width $\tw$ and height
bounded by $\OO(\tw \cdot \log |V(G)|)$.
Also, we assume that $G$ contains all edges connecting pairs $u, v \in X_t$ for
each node $t$.
Moreover, we are given an integer~$r$ bounded by $\OO((1/\epsilon)|V(G)|)$.
Our goal is to design a dynamic programming algorithm that computes the minimum
number of hubs that satisfy each demand with a path of length~$r$.
The overall idea is similar to that of the algorithm for \kcfull by
Demaine~\etal~\cite{DemaineFHT05}, except that we consider a tree decomposition,
instead of a branch decomposition, and that the computed solution will satisfy
demands only approximately.

Consider some fixed global optimal solution $H^*$ and a node $t$ of the tree
decomposition. Let us discuss possible candidates for a subproblem definition.
The subgraph $G_t$ corresponding to $t$ in the decomposition contains a subset
of $H^*$ that satisfies a subset $\De_t$ of the demands.
The shortest path serving each demand with a hub of $H^* \cap V(G_t)$ is either
completely contained in $G_t$, or it must cross some vertex of the bag~$X_t$.
Thus, as in~\cite{DemaineFHT05}, we guess the distance $i$ from each vertex $u$
in $X_t$ to the closest hub in $H^*$, and assign ``color'' $\downa{i}$ to $u$ to
mean that the corresponding shortest path is in $G_t$, and color $\upa{i}$ to
mean otherwise.

Since the number of demands may be large, we cannot include $\De_t$ as part of
the subproblem definition.
For \kcfull, if the shortest path serving a vertex in $G_t$ crosses a vertex $u
\in X_t$, then the length of this path can be bounded locally using the color
of~$u$, and the subproblem definition may require serving all vertices.
For \makhc, however, there might be demands $(a,b)$ such that~$a$~is in~$G_t$,
while~$b$~is not, thus the coloring of $X_t$ is not sufficient to bound the
length of a path serving~$(a,b)$.

Instead of guessing $\De_t$, for each coloring $c$ of $X_t$, we require that
only a subset~$D_t(c)$ must be satisfied in the subproblem, and they can be
satisfied by a path of length at most $2r$.
Later, we show that the other demands in $\De_t$ are already satisfied by the
hubs corresponding to the coloring of $X_t$.
More specifically, we would like to compute $A_t(c)$ as the minimum number of
hubs in $G_t$ that satisfy each demand in~$D_t(c)$ with a path of length at most
$2r$ and that respect the distances given by~$c$.

Since we preprocessed the graph in Section~\ref{section:preliminaries}, there
must be a hub in $H^*$ to each vertex of $X_t$ at distance at most $r$.
Thus, the number of distinct colorings to consider for each $t$ is bounded by
$r^{\OO(\tw)}$.
To get an algorithm parameterized only by~$\tw$, we need one more ingredient:
in the following, the value of each color is stored approximately as an integer
power of $(1+\delta)$, for some $\delta > 0$. Later, using the framework of
approximate addition trees, for any constant $\epsilon > 0$, we can set $\delta$
such that the number of subproblems is bounded by
$\OO^*((\tw/\epsilon)^{\OO(\tw)})$, and demands are satisfied by a path of
length at most $(1+\epsilon)2r$.

The set of approximate colors is
\begin{align*}
  \Sigma =
  \{ \downa{0}\}
  \cup
  \{ \,
  \upa{i}, \, \downa{i} \; :  \;
  j \in\integersnn \, , \,
  i = (1+\delta)^j \, , \,
  i \le (1 + \epsilon)r
  \, \}.
\end{align*}
A coloring of $X_t$ is represented by a function $c : X_t \to \Sigma$. For each
coloring~$c$, we compute a set of demands that are ``satisfied'' by $c$.

\begin{definition}
  Define $S_t(c)$ as the set of demands $(a,b)$ for which there exists ${u \in
      X_t}$ with ${c(u) \in \{\upa{i},\downa{i}\}}$ and such that ${d(a,u) + 2i +
      d(u,b) \leq (1+\epsilon)2r}$.
\end{definition}

The intuition is that a demand $(a, b) \in S_t(c)$ can be satisfied by a hub
close to~$u$ by a path of length at most $(1+\epsilon)2r$.
Also, we compute a set of demands that must be served by a hub in~$G_t$ by the
global optimal solution.

\begin{definition}
  Define $D_t(c)$  as the set of demands $(a,b)$ such that ${(a,b) \notin
      S_t(c)}$ and either: \,
  \begin{enumerate*}[$(i)$]
    \item $a,b \in V(G_t)$;  \, or \,
    \item $a \in V(G_t)$, $b \notin V(G_t)$ and there is $h \in V(G_{ab})
    \cap V(G_t)$ such that $d(h, V(G_{ab})\cap X_t) > r/2$.
  \end{enumerate*}
\end{definition}

We will show in Lemmas~\ref{lemma:d_in_td}~and~\ref{lemma:td_in_d_cup_s} that
$D_t(c) \subseteq \De_t \subseteq D_t(c) \cup S_t(c)$, thus we only need to take
care of demands in $D_t(c)$ in the subproblem.
Formally, for each node~$t$ of the tree decomposition  and coloring $c$ of
$X_t$, our algorithm computes a number $A_t(c)$ and a set of hubs $H \subseteq
\hubs \cap V(G_t)$ of size~$A_t(c)$ that satisfies the conditions below.
\begin{enumerate}

  \item[\cone] For every $u \in X_t$, if $c(u) = \downa{i}$, then there exists
      $h \in H$ and a shortest path $P$ from~$u$~to~$h$ of length at most~$i$
      such that $V(P) \subseteq V(G_t)$;

  \item[\ctwo] For every $(a,b) \in D_t(c)$, $\min_{h \in H} d(a,h) + d(h,b)
      \leq (1+\epsilon)2r$.

\end{enumerate}
If the algorithm does not find one such set, then it assigns $A_t(c) = \infty$.
We describe next how to compute $A_t(c)$ for each node type.

For a \emph{leaf node} $t$, we have $V(G_t) = \emptyset$, then $H = \emptyset$
satisfies the conditions, and we set $A_t(\cempty) = 0$, where $\cempty$ denotes
the empty coloring.

For an \emph{introduce node} $t$ with child $t'$, let $u$ be the introduced
vertex, such that $X_t = X_{t'} \cup \{u\}$.
Let $I_t(c)$ be the set of colorings $c'$ of $X_{t'}$ such that $c'$ is the
restriction of $c$ to $X_{t'}$ and, if $c(u) = \downa{i}$ for some $i > 0$,
there is ${v \in X_{t'}}$ with $c'(v) = \downa{j}$ such that $i = d(u,v) \oplus
j$.
Note that this set is either a singleton or is empty.
If $I_t(c)$ is empty, discard $c$. Define:
\begin{align*}
  A_t(c) =
  \min_{\substack{c' \in I_t(c): \\D_t(c) \subseteq D_{t'}(c')}}
  \begin{cases}
    A_{t'}(c') + 1 & \text{if } c(u) = \downa{0}, \\
    A_{t'}(c')     & \text{otherwise}.
  \end{cases}
\end{align*}
If $H'$ is the solution corresponding to $A_{t'}(c')$, we
output $H = H' \cup \{u\}$ if $c(u) = \downa{0}$, or $H = H'$ otherwise.

For a \emph{forget node} $t$ with child $t'$, let $u$ be the forgotten vertex,
such that ${X_t = X_{t'} \setminus \{u\}}$.
Let $F_t(c)$ be the set of colorings $c'$ of $X_{t'}$ such that $c$ is the
restriction of $c'$ to $X_t$ and, if~$c'(u) = \upa{i}$, then there is $v \in
X_t$ such that $c(v) = \upa{j}$ and $i = d(u,v) \oplus j$.
If $F_t(c)$ is empty, discard $c$. Define:
\begin{align*}
  A_t(c) = \min_{\substack{c' \in F_t(c): \\ D_t(c) \subseteq D_{t'}(c') \cup
  S_{t'}(c')}} A_{t'}(c').
\end{align*}
We output as solution the set $H = H'$, where $H'$ corresponds to the solution
of the selected subproblem in $t'$.

For a \emph{join node} $t$ with children $t'$ and $t''$, we have $X_t = X_{t'} =
X_{t''}$.
Let~$J_t(c)$ be the set of pairs of colorings $(c',c'')$ of $X_t$ such that, for
every $u \in X_t$, when $c(u)$ is~$\downa{0}$~or~$\upa{i}$, then ${c'(u) =
c''(u) = c(u)}$;
else, if $c(u)$ is $\downa{i}$, then $(c'(u),c''(u))$ is
either~$(\upa{i},\downa{i})$ or~$(\downa{i},\upa{i})$.
If $J_t(c)$ is empty, discard $c$. Define:
\begin{align*}
  A_t(c) = \min_{\substack{(c',c'') \in J_t(c): \\D_t(c) \subseteq D_{t'}(c')
  \cup D_{t''}(c'')}} A_{t'}(c') + A_{t''}(c'') - h(c),
\end{align*}
where $h(c)$ is the number of vertices $u$ in $X_t$ such that $c(u) =
\downa{0}$. We output a solution $H = H' \cup H''$, where $H'$ and $H''$ are
the solutions corresponding to $t'$~and~$t''$, respectively.

In the next lemma, we show that the algorithm indeed produces a solution of
bounded size that satisfies both conditions.

\begin{lemma}
  \label{lemma:algorithm_outputs_solution}
  If $A_t(c) \neq \infty$, then the algorithm outputs a set $H \subseteq \hubs
  \cap V(G_t)$, with $|H| \leq A_t(c)$, that satisfies \cone and \ctwo.
\end{lemma}

\begin{proof}
We prove the lemma by induction on the height of node $t$, thus assume the lemma
holds for nodes below $t$.
For \emph{leaves}, the algorithm outputs an empty set, satisfying both
conditions.

For an \emph{introduce node} $t$ with child $t'$ and $u \in X_t \setminus
X_{t'}$, let $H'$ be the solution corresponding to $t'$ with coloring $c'$.
Since $c'$ is the restriction of $c$ to~$X_{t'}$, condition~\cone is satisfied
for every $v \in X_{t'}$, by induction. If ${c(u) = \downa{0}}$, then it is
satisfied for $u$, since, in this case, $u \in H$. Else, if $c(u) = \downa{i}$
for $i > 0$, then it is also satisfied, since in this case there is $v \in
X_{t'}$ with $c'(v) = \downa{j}$ such that $i = d(u,v) \oplus j$.
Condition \ctwo is satisfied as well, since $D_t(c) \subseteq D_{t'}(c')$, and
$H'$ satisfies \ctwo.

For a \emph{forget node} $t$ with child $t'$ and $u \in X_{t'} \setminus X_t$,
we have that ${H = H'}$, where $H'$ is the solution to $t'$ corresponding to
some coloring $c'$ of $X_{t'}$.
Since $c$~is the restriction of $c'$ to $X_t$, $H$ satisfies \cone by
induction.
For~\ctwo, let ${(a,b) \in D_t(c)}$, and remember that ${D_t(c) \subseteq
    D_{t'}(c') \cup S_{t'}(c')}$.
If $(a,b) \in D_{t'}(c')$, then this demand is satisfied by $H$ with cost at
most $(1+\epsilon)2r$. Else, $(a,b) \in S_{t'}(c')$,  but $(a,b) \notin S_t(c)$.
Thus, for the forgotten vertex $u$, we have $c'(u) \in \{ \upa{i},\downa{i}\}$
and $d(a,u) + 2i + d(u,b) \leq (1+\epsilon)2r$. We consider two cases:
\begin{itemize}

  \item If $c'(u) = \downa{i}$, then, since $H$ satisfies \cone, there is $h
    \in H$ such that the distance from $u$ to $h$ is at most $i$. Thus
    condition~\ctwo is satisfied, because
    \begin{align*}
      d(a,h) + d(h,b) & \leq d(a,u) + 2i + d(u,b) \leq (1+\epsilon)2r.
    \end{align*}

  \item If $c'(u) = \upa{i}$, there is $v \in X_t$ with ${c(v) = \upa{j}}$ and
    ${i = d(u,v) \oplus j}$.
    We get
    \begin{align*}
      d(a,v) + 2j + d(v,b)
       & \leq d(a,u) + 2(d(u,v) + j) + d(u,b) \\
       & \leq d(a,u) + 2i + d(u,b)            \\
       & \leq (1+\epsilon)2r,
    \end{align*}
    where we used ${d(u,v) + j \leq d(u,v) \oplus j = i}$ in the second inequality.
    But this means that ${(a,b) \in S_t(c)}$, which is a contradiction.

\end{itemize}

For a \emph{join node} $t$ with children $t'$ and $t''$, let $H'$ and $H''$ be
solutions for the subproblems at $t'$ and $t''$ corresponding to the selected
pair of colorings $c'$~and~$c''$.
We claim that $H = H' \cup H''$ satisfies both conditions. For \cone, note that
if $c(u) = \downa{i}$, for some $u \in X_t$, then $c'(u) = \downa{i}$ or $c''(u)
= \downa{i}$. For \ctwo, note that, for $(a,b) \in D_t(c)$, we have $(a,b) \in
D_{t'}(c')\cup D_{t''}(c'')$, and thus this demand is satisfied with cost at
most $(1+\epsilon)2r$ by a vertex in $H'$ or $H''$.\qed
\end{proof}

Let $\troot$ be the root of the tree decomposition and $\cempty$ be the empty
coloring. Since the bag corresponding to the root node is empty, we have
$S_{\troot}(\cempty) = \emptyset$ and thus $D_{\troot}(\cempty) = \demands$.
Therefore, if $A_{\troot}(\cempty) \leq k$,
Lemma~\ref{lemma:algorithm_outputs_solution} implies that the set of hubs~$H$
computed by the algorithm is a feasible solution that satisfies each demand with
cost at most $(1+\epsilon)2r$.
In the next section, we bound the size of $H$ by the size of the global optimal
solution~$H^*$.

%% file: content/analysis.tex
\section{Analysis}
\label{section:analysis}

For each node $t$ of the tree decomposition, we want to show that the number of
hubs computed by the algorithm for some coloring $c$ of $X_t$ is not larger than
the number of hubs of~$H^*$ contained in $G_t$, that is, we would like to show
that $A_t(c) \le |H^* \cap V(G_t)|$ for some~$c$.
If the distances from each vertex $u \in X_t$ to its closest hub in~$H^*$ were
stored exactly, then the partial solution corresponding to~$H^*$ would induce
one such coloring $\ce_t$, and we could show the inequality for this particular
coloring.
More precisely, for each $u \in V(G)$, let $h^*(u)$ be a hub of $H^*$ such that
$d(u, h^*(u))$ is minimum and $P^*(u)$ be a corresponding shortest path.
Assume that each $P^*(u)$ is obtained from a shortest path tree to $h^*(u)$ and
that it has the minimum number of edges among the shortest paths.
The \emph{signature} of $H^*$ corresponding to a partial solution in $G_t$ is a
function $\ce_t$ on $X_t$ such that
\begin{align*}
  \ce_t(u) =
  \begin{cases}
    \downa{d(u, h^*(u))} & \text{if } V(P^*(u)) \subseteq V(G_t) , \\
    \upa{d(u, h^*(u))}   & \text{otherwise}.
  \end{cases}
\end{align*}

Since distances are stored approximately as integer powers of $(1 + \delta)$,
the function~$\ce_t$ might not be a valid coloring.
Instead, we show that the algorithm considers a coloring $\ca_t$ with roughly
the same values of~$\ce_t$ and that its values are computed by approximate
addition trees.
We say that an addition tree and an approximate addition tree are
\emph{corresponding} if they are isomorphic and have the same input values.
Also, recall that a coloring $c$ of $X_t$ is \emph{discarded} by the algorithm
if the set $I_t(c)$, $F_t(c)$ or $J_t(c)$ corresponding to $t$ is empty.

\begin{lemma}
  \label{lemma:correspondence}
  Let $\ell_{\troot}$ be the height of the tree decomposition.
  There exists a coloring~$\ca_t$ that is not discarded by the algorithm and
  such that, for every $u \in X_t$, the values $\ce_t(u)$~and~$\ca_t(u)$ are
  computed, respectively, by an addition tree and a corresponding approximate
  addition tree of height at most~$2\ell_{\troot}$.
\end{lemma}

\begin{proof}
  A \emph{partial addition tree} is a pair $(T, p)$, where $T$ is an addition
  tree and $p$ is a leaf of $T$.
  The vertex $p$ represents a subtree that computes a pending value~$x_p$, and
  may be replaced by some other (partial) addition tree that computes this value.

  For some node $t$, let $\ell_t$ be the height of $t$ and define $U_t$ as the
  set of vertices ${u \in X_t}$ such that $\ce_t(u) = \upa{i}$ for some $i$.
  We say that a vertex ${v \in V(G_t) \setminus U_t}$
  is~\emph{\mbox{$t$-complete}} according to the following cases:
  \begin{itemize}

    \item if $V(P^*(v)) \subseteq V(G_t)$ and $v \in X_t$, then $d(v,h^*(v))$ is
      computed by an addition tree of height at most~$\ell_t$;

    \item if $V(P^*(v)) \subseteq V(G_t)$ and $v \notin X_t$, then $d(v,h^*(v))$ is
      computed by an addition tree of height at most~$2\ell_t$;

    \item if $V(P^*(v)) \not\subseteq V(G_t)$, then $d(v,h^*(v))$ is computed by
      a partial addition tree $(T,p)$ of height at most~$\ell_t$ such that ${x_p
          = d(w, h^*(w))}$ for some $w \in U_t$.

  \end{itemize}
  We will show by induction on the height of $t$ that every ${v \in V(G_t)
      \setminus U_t}$ is \mbox{$t$-complete}.
  The claim holds trivially for leaves, thus suppose that $t$ is not a leaf.

  Assume $t$ is an \emph{introduce node} with child $t'$, and let $u$ be the
  introduced vertex.
  Since ${U_{t'} \subseteq U_t}$, if $v \in V(G_{t'}) \setminus U_{t'}$, then
  $v$ is $t$-complete by the induction hypothesis.
  Else, we have ${v = u}$ and, since $v \notin U_t$, $\ce_t(v) = \downa{d(v,
  h^*(v))}$. Thus, there is ${w \in X_{t'} \setminus U_{t'}}$ such that
  ${\ce_{t'}(w) = \downa{d(w, h^*(w))}}$ and $d(v, h^*(v)) = d(v, w) + d(w,
  h^*(w))$.
  Since $d(w, h^*(w))$ can be computed by an addition tree of height at most
  $\ell_{t'}$, this implies that $d(v, h^*(v))$ can be computed by an addition
  tree of height at most ${\ell_{t'} + 1 \le \ell_t}$.

  Now, assume $t$ is a \emph{forget node} with child $t'$, and let $u$ be the forgotten
  vertex.
  Since ${V(G_t) = V(G_{t'})}$, if $V(P^*(v)) \subseteq V(G_{t'})$, then $v$ is
  $t$-complete by the induction hypothesis.
  Otherwise, by the induction hypothesis, $d(v,h^*(v))$ is computed by a partial
  addition tree $(T,p)$ of height at most~$\ell_{t'}$ such that ${x_p = d(w',
  h^*(w'))}$ for some $w' \in U_{t'}$.
  If $w' \in U_t$, then $v$ is $t$-complete. So, assume $w' \in U_{t'} \setminus
  U_t$, which implies that $w'$ is the forgotten vertex~$u$ and $\ce_{t'}(u) =
  \upa{d(u, h^*(u))}$.
  Thus, $P^*(u)$ crosses some vertex $w \in U_t$ such that $d(u, h^*(u)) = d(u,
  w) + d(w, h^*(w))$.
  It follows that $d(u, h^*(u))$ can be computed by a partial addition tree
  $(T_u, p_u)$ of height~$1$ such that $x_{p_u} = d(w, h^*(w))$.
  Therefore, we can replace the vertex $p$ by the subtree~$T_u$, and the height
  of~$T$ becomes at most $\ell_{t'} + 1 \le \ell_t$.

  Finally, assume $t$ is a \emph{join node} with children $t'$ and $t''$, and
  recall that $X_t = X_{t'} = X_{t''}$.
  If $v \in X_t$, then $V(P^*(v)) \subseteq V(G_{t'})$ or $V(P^*(v)) \subseteq
  V(G_{t''})$, because $X_t$ induces a clique and $P^*(v)$ is a shortest path
  with minimum number of edges.
  Thus, $v$ is $t$-complete by the induction hypothesis.
  Otherwise, $v \in V(G_t) \setminus X_t$. Assume $v \in V(G_{t'}) \setminus
  X_t$, as the other case is analogous.
  By the induction hypothesis for~$t'$, $d(v,h^*(v))$ is computed by a partial
  addition tree $(T',p)$ of height at most~$\ell_{t'}$ such that ${x_p = d(w,
  h^*(w))}$ for some $w \in U_{t'}$.
  If $w \in U_t$, then $v$ is $t$-complete. Thus, assume $w \notin U_t$, which
  implies that $V(P^*(w)) \subseteq V(G_t)$.
  Again, since $X_t$ induces a clique, $P^*(w)$ is included in $V(G_{t'})$ or
  $V(G_{t''})$, but since $w \in U_{t'}$, we have $V(P^*(w)) \subseteq
  V(G_{t''})$.
  It follows that $\ce_{t''}(w) =\downa{d(w, h^*(w))}$.
  By the induction hypothesis for $t''$, $d(w,h^*(w))$ is computed by an
  addition tree $T''$ of height at most~$\ell_{t''}$.
  Therefore, we can replace the vertex $p$ by the subtree $T''$, and the height
  of $T'$ becomes at most $\ell_{t'} + \ell_{t''} \le 2 \ell_t$.
  This completes the induction.

  \medskip

  For the root node $\troot$, we have $X_{\troot} = \emptyset$, thus for every
  $v \in V(G)$, the distance $d(v,h^*(v))$ is computed by an addition tree $T_v$
  of height at most~$2\ell_{\troot}$.
  Let $\bar{T}_v$ be the approximate addition tree corresponding to $T_v$, and
  define $\bar{d}(v)$ as the output of~$\bar{T}_v$.
  For every node~$t$, and $u \in X_t$, if $\ce_t(u) = \downa{d(u, h^*(u))}$,
  define $\ca_t(u) = \downa{\bar{d}(u)}$; else, define $\ca_t(u) =
  \upa{\bar{d}(u)}$.
  By repeating the arguments above, and replacing the addition operator by
  $\oplus$, one can show that, for every $t$, the coloring $\ca_t$ is not
  discarded by the algorithm.\qed
\end{proof}

By setting $\delta = \epsilon/(2\ell_{\troot} + 1)$,
Theorem~\ref{theorem:lampis} implies the next lemma.

\begin{lemma}
    \label{lemma:bound_ca}
    For every $u \in X_t$, if $\ce_t(u) \in \{\upa{i}, \downa{i}\}$ and
    $\ca_t(u) \in \{\upa{j}, \downa{j}\}$, then $j \le (1+ \epsilon) i$.
\end{lemma}

Recall that~$H^*$ is a fixed global optimal solution that satisfies each demand
with cost~$r$.
Our goal is to bound $A_t(\ca_t) \le |H^* \cap V(G_t)|$ for every node~$t$, thus
we would like to determine the subset of demands $\De_t$ that are necessarily
satisfied by hubs $H^* \cap V(G_t)$ in the subproblem definition.
This is made precise in the following.

\begin{definition}
  \label{def-d-star}
  $\De_t = \{(a,b) \in \demands:
  \min_{h \in H^*\setminus V(G_t)} d(a,h)
  + d (h,b) > r\}$.
\end{definition}

Since the algorithm cannot determine $\De_t$, we show that, for each node $t$,
it outputs a solution~$H$ for the subproblem corresponding to $A_t(\ca_t)$ that
satisfies every demand in $D_t(\ca_t)$.
In Lemma~\ref{lemma:d_in_td}, we show that every demand in $D_t(\ca_t)$ is also
in $\De_t$, as, otherwise, there could be no solution with size bounded by $|H^*
\cap V(G_t)|$.
Conversely, we show in Lemma~\ref{lemma:td_in_d_cup_s} that a demand in $\De_t$
that is not in $D_t(\ca_t)$ must be in $S_t(\ca_t)$, thus all demands are
satisfied.

\begin{lemma}
  \label{lemma:d_in_td}
  $D_t(\ca_t) \subseteq \De_t$.
\end{lemma}
\begin{proof}
  Let ${(a,b) \in D_t(\ca_t)}$ and consider an arbitrary hub ${h^* \in H^*}$
  that satisfies~$(a,b)$ with cost~$r$. We will show that ${h^* \in V(G_t)}$,
  and thus ${(a,b) \in \De_t}$.
  For the sake of contradiction, assume that $h^* \in V(G) \setminus V(G_t)$.

  First we claim that $d(h^*, V(G_{ab}) \cap X_t) > r/2$. If not, then let $u
  \in V(G_{ab}) \cap X_t$ be a vertex with $\ca_t(u) \in \{\upa{i},\downa{i}\}$
  such that $d(u,h^*) \leq r/2$.
  Because the closest hub to~$u$ has distance at least~$i/(1+\epsilon)$, we have
  $i \leq (1+\epsilon)d(u,h^*) \leq (1+\epsilon)r/2$, but since $u \in
  V(G_{ab})$, this implies that $(a,b) \in S_t(\ca_t)$, and thus $(a, b) \notin
  D_t(\ca_t)$.
  Then, it follows that indeed $d(h^*, V(G_{ab}) \cap X_t) > r/2$.

  Now we show that it cannot be the case that $a, b \in V(G_t)$. Suppose that
  $a, b \in V(G_t)$. Consider the shortest path from $a$ to $h^*$, and let $u$
  be the last vertex of this path that is in $V(G_t)$.
  Since $X_t$ separates $V(G_t) \setminus X_t$ from $V(G) \setminus V(G_t)$, it
  follows that $u \in X_t$. From the previous claim, $d(h^*, u) > r/2$, and thus
  $d(h^*, a) > r/2$. Analogously, $d(h^*, b) > r/2$, but then $d(a, h^*) +
  d(h^*, b) > r$, which contradicts the fact that $h^*$ satisfies $(a,b)$ with
  cost~$r$.
  This contradiction comes from supposing that $a, b \in V(G_t)$. Thus, either
  $a$ or $b$ is not in $V(G_t)$.

  Assume without loss of generality that $a \in V(G_t)$ and $b \notin V(G_t)$.
  From the definition of $D_t(\ca_t)$, we know that there exists $h \in V(G_{ab})
  \cap V(G_t)$ such that $d(h, V(G_{ab}) \cap X_t) > r/2$. Let $P$ be a path
  from $a$~to~$b$ crossing $h^*$ with length at most $r$. Similarly, since $h
  \in V(G_{ab})$, there exists a path $Q$ from $a$~to~$b$ crossing $h$ with
  length at most $r$.
  Let $u$ be the last vertex of $P$ with $u \in X_t$, and let $v$ be the last
  vertex of $Q$ with $v \in X_t$ (see Figure~\ref{figure:closed_walk}).
  Concatenating $P$ and $Q$ leads to a closed walk of length at most $2r$. This
  walk crosses $u$, $h^*$, $v$ and $h$, and thus
  \begin{align}
    2r & \geq d(a,h^*) + d(h^*,b) + d(a,h) + d(h,b)\nonumber \\
       & = d(u,h^*) + d(h^*,v) + d(v,h) + d(h,u)\label{pointer1} \\
       & > 2r,\nonumber
  \end{align}
  where we used the fact that each term in (\ref{pointer1}) is greater than
  $r/2$. This is a contradiction, so $h^* \in V(G_t)$ and then $(a,b) \in
  \De_t$.\qed

  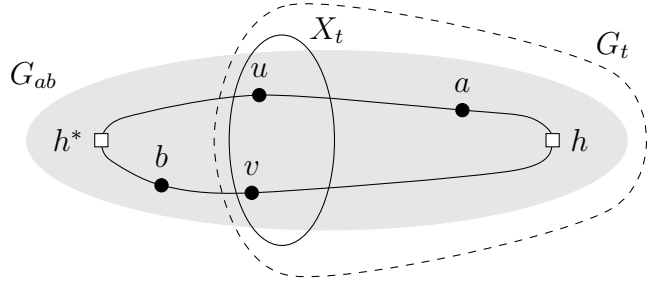
\begin{figure}[t]
    \centering
    \begin{tikzpicture}
      \coordinate (coord_hs) at (0.0, 0.0);
      \coordinate (coord_h) at (6.0, 0.0);
      \coordinate (coord_u) at (2.1, 0.6);
      \coordinate (coord_v) at (2.0, -0.7);
      \coordinate (coord_a) at (4.8, 0.4);
      \coordinate (coord_b) at (0.8, -0.6);
      \fill[black!10!white] (3,0) ellipse (4cm and 1.2cm);
      \node[label={$G_{ab}$}] at (-0.9, 0.4) {};
      \draw (2.4,0) ellipse (0.7cm and 1.4cm);
      \node[label={$X_t$}] at (3.0, 1) {};

      \draw plot [smooth] coordinates {(coord_hs) (0.3,0.3) (coord_u)
          (coord_a) (5.7,0.3) (coord_h) (5.5,-0.4) (coord_v) (coord_b)
          (0.1,-0.25) (coord_hs)};
      \draw[dashed] plot [smooth cycle] coordinates {(1.5,0) (2.5,1.8)
          (6.8,0.9) (6.8,-0.9) (2.5,-1.8) };
      \node[label={$G_t$}] at (6.8, 0.8) {};
      \node[hubwhite,label=left:{$h^*$}] at (coord_hs) (hs) {};
      \node[hubwhite,label=right:{$h$}] at (coord_h) (h) {};
      \node[vertexblack,label=above:{$u$}] at (coord_u) (u) {};
      \node[vertexblack,label=above:{$v$}] at (coord_v) (v) {};
      \node[vertexblack,label=above:{$a$}] at (coord_a) (a) {};
      \node[vertexblack,label=above:{$b$}] at (coord_b) (b) {};
    \end{tikzpicture}
    \caption{Closed walk formed by $P$ and $Q$.}
    \label{figure:closed_walk}
  \end{figure}

\end{proof}

\begin{lemma}
  \label{lemma:td_in_d_cup_s}
  $\De_t \subseteq D_t(\ca_t) \cup S_t(\ca_t)$.
\end{lemma}
\begin{proof}
  Let $(a,b) \in \De_t$. Assume $(a,b) \notin S_t(\ca_t)$, as otherwise we are
  done. If $a,b \in V(G_t)$, we have ${(a,b) \in D_t(\ca_t)}$. Thus, suppose
  without loss of generality that ${a \in V(G_t)}$ and ${b \notin V(G_t)}$.
  Since $(a,b) \in \De_t$, there is $h^* \in H^* \cap V(G_{ab}) \cap V(G_t)$.
  Let $u \in V(G_{ab}) \cap X_t$ with $\ca_t(u) \in \{\upa{i}, \downa{i}\}$
  for some~$i$.
  Because $(a,b) \notin S_t(\ca_t)$, we have ${i > (1 + \epsilon) r/2}$. But the
  distance from $u$ to the closest hub in $H^*$ is at least ${i /
  (1+\epsilon)}$, thus ${ i \leq (1+\epsilon) d(u,h^*)}$. It follows that
  ${d(u,h^*) > r/2}$.
  Therefore, $h^* \in V(G_{ab}) \cap V(G_t)$ and $d(h^*, V(G_{ab}) \cap X_t) >
  r/2$, and then $(a,b) \in D_t(\ca_t)$.\qed
\end{proof}

Before bounding the number of hubs opened by the algorithm, we prove some
auxiliary results.

\begin{lemma}
  \label{lemma:condition_introduce}
  If $t$ is an introduce node with child $t'$, then ${D_t(\ca_t) \subseteq
      D_{t'}(\ca_{t'})}$.
\end{lemma}
\begin{proof}
  We claim that $\De_t \setminus \De_{t'} \subseteq S_t(\ca_t)$.
  Let $u$ be introduced vertex, and note that $V(G_t) \setminus V(G_{t'}) =
  \{u\}$.
  If $(a, b) \in \De_t \setminus \De_{t'}$, by definition, we know that
  ${\min_{h \in H^*\setminus V(G_t)} d(a,h) + d (h,b) > r}$, but ${\min_{h \in
  H^*\setminus V(G_{t'})} d(a,h) + d (h,b) \le r}$.
  This can only happen if $u \in H^*$, so $\ca_t(u) = \downa{0}$, and then $(a,
  b) \in S_t(\ca_t)$.

  Since $\De_{t'} \subseteq \De_t$ and $S_{t'}(\ca_{t'}) \subseteq S_t(\ca_t)$,
  the claim implies $\De_t \setminus S_t(\ca_t) \subseteq \De_{t'} \setminus
  S_{t'}(\ca_{t'})$.
  Using Lemmas~\ref{lemma:d_in_td}~and~\ref{lemma:td_in_d_cup_s}, we get
  \begin{align*}
    D_t(\ca_t)
    \subseteq
    \De_t     \setminus S_t(\ca_t)
    \subseteq
    \De_{t'}  \setminus S_{t'}(\ca_{t'})
    \subseteq
    D_{t'}(\ca_{t'}).\tag*{\qed}
  \end{align*}
\end{proof}

\begin{lemma}
  \label{lemma:condition_forget}
  If $t$ is a forget node with child $t'$, then $D_t(\ca_t) \subseteq
  D_{t'}(\ca_{t'}) \cup S_{t'}(\ca_{t'})$.
\end{lemma}
\begin{proof}
  In this case $V(G_t) = V(G_{t'})$, thus $ \De_t = \De_{t'}$. Using
  Lemmas~\ref{lemma:d_in_td}~and~\ref{lemma:td_in_d_cup_s},
  \begin{align*}
    D_t(\ca_t)
    \subseteq
    \De_t =  \De_{t'}
    \subseteq
    D_{t'}(\ca_{t'}) \cup S_{t'}(\ca_{t'}).\tag*{\qed}
  \end{align*}
\end{proof}

\begin{lemma}
  \label{lemma:condition_join}
  If $t$ is a join node with children $t'$ and $t''$, then $D_t(\ca_t) \subseteq
  D_{t'}(\ca_{t'}) \cup D_{t''}(\ca_{t''})$.
\end{lemma}
\begin{proof}
  We claim that $\De_t \setminus (\De_{t'} \cup \De_{t''}) \subseteq S_t(\ca_t)$.
  Let $(a, b) \in \De_t \setminus (\De_{t'} \cup \De_{t''})$ and suppose,
  for a contradiction, that $(a, b) \notin S_t(\ca_t)$.
  Then, for every $h^* \in H^*$ and $u \in V(G_{ab}) \cap X_t$, we have $d(h^*,
  u) > r/2$.
  Since $(a,b) \in \De_t$, but $(a, b) \notin \De_{t''}$, there is $h' \in H^*
  \cap V(G_{t'}) \setminus X_t$ that satisfies $(a,b)$.
  Similarly, there is $h'' \in H^* \cap V(G_{t''}) \setminus X_t$.
  Now, $h', h''\in V(G_{ab})$, but the diameter of $G_{ab}$ is at most~$r$, thus
  $d(h', h'') \le r$.
  Since $X_t$ separates $h'$ and $h''$, there is $u \in X_t$ with $d(h', u) +
  d(u, h'') \le r$. Thus, either $d(h', u) \le r/2$ or $d(h'', u) \le r/2$, a
  contradiction.
  This implies $(a,b) \in S_t(\ca_t)$.

  Observe that $\De_{t'} \cup \De_{t''} \subseteq \De_t$ and $S_t(\ca_t) =
  S_{t'}(\ca_{t'}) = S_{t''}(\ca_{t''})$.
  Combining with Lemmas~\ref{lemma:d_in_td}~and~\ref{lemma:td_in_d_cup_s}, we get
  \begin{align*}
    D_t(\ca_t)
    \subseteq
    \De_t     \setminus S_t(\ca_t)
    \subseteq
    (\De_{t'} \setminus S_{t'}(\ca_{t'})) \cup (\De_{t''} \setminus S_{t''}(\ca_{t''}))
    \subseteq
    D_{t'}(\ca_{t'}) \cup D_{t''}(\ca_{t''}).\tag*{\qed}
  \end{align*}
\end{proof}

Combining Lemma~\ref{lemma:correspondence} and
Lemmas~\ref{lemma:condition_introduce}--\ref{lemma:condition_join}, we can show
that the algorithm does not open too many hubs.

\begin{lemma}
  \label{lemma:not_many_hubs}
  $A_t(\ca_t) \leq |H^* \cap V(G_t)|$.
\end{lemma}

\begin{proof}
  Assume the lemma holds for the children of~$t$.
  For a \emph{leaf node}, the output set is empty, and the inequality is
  satisfied trivially.

  Let $t$ be an \emph{introduce node} with child $t'$ and $u \in X_t \setminus
  X_{t'}$.
  From Lemmas~\ref{lemma:correspondence}~and~\ref{lemma:condition_introduce}, we
  know that $\ca_{t'} \in I_t(\ca_t)$ and $D_t(\ca_t) \subseteq
  D_{t'}(\ca_{t'})$.
  Thus, if $\ca_t(u) = \downa{0}$, we have $u \in H^*$ and $A_t(\ca_t) =
  A_{t'}(\ca_{t'}) + 1 \leq |H^* \cap V(G_{t'})| + 1 = |H^* \cap V(G_t)|$.
  Otherwise, $A_t(\ca_t) = A_{t'}(\ca_{t'}) \leq |H^* \cap V(G_{t'})| = |H^*
  \cap V(G_t)|$.

  Let $t$ be a \emph{forget node} with child $t'$ and $u \in X_{t'} \setminus
  X_t$.
  From Lemmas~\ref{lemma:correspondence}~and~\ref{lemma:condition_forget}, we
  know that $\ca_{t'} \in F_t(\ca_t)$ and $D_t(\ca_t) \subseteq D_{t'}(\ca_{t'})
  \cup S_{t'}(\ca_{t'})$.
  Thus, $A_t(\ca_t) \leq A_{t'}(\ca_{t'}) \leq |H^* \cap V(G_{t'})| = |H^* \cap
  V(G_t)|$.

  Let $t$ be a \emph{join node} with children $t'$ and $t''$.
  From Lemmas~\ref{lemma:correspondence}~and~\ref{lemma:condition_join}, we know
  that $(\ca_{t'}, \ca_{t''}) \in J_t(\ca_t)$ and $D_t(\ca_t) \subseteq
  D_{t'}(\ca_{t'}) \cup D_{t''}(\ca_{t''})$.
  Let $H'$ and $H''$ be the output solutions corresponding to $t'$ and $t''$,
  respectively.
  We have
  \begin{align*}
    |H| & = |H'| + |H''| - |H' \cap H''|                               \\
        & \leq A_{t'}(\ca_{t'}) + A_{t''}(\ca_{t''}) - h(\ca_t)        \\
        & \leq |H^* \cap V(G_{t'})| + |H^* \cap V(G_{t''})| - h(\ca_t) \\
        & = |H^* \cap V(G_t)|.\tag*{\qed}
  \end{align*}
\end{proof}

Now we can state the main result.

\begin{theorem}
    \label{theorem:main}
    For every $\epsilon > 0$, there is a parameterized
    $(2+\epsilon)$-approximation algorithm for \makhc running in time $\OA((\tw
    / \epsilon)^{\OO(\tw)})$.
\end{theorem}

\begin{proof}
  Consider a preprocessed instance $(G,\clients,\hubs,\demands,k)$ of \makhc, in
  which the optimal value \opt is an integer bounded by
  $\OO(\frac{1}{\epsilon}|V(G)|)$.
  We run the dynamic programming algorithm for each $r = 1, 2, \dots$, and
  output the first solution with no more than~$k$ hubs.
  Next, we show that the dynamic programming algorithm either correctly decides
  that there is no solution of cost $r$ that opens $k$ hubs, or finds a solution
  of cost $(1+\epsilon)2r$ that opens $k$ hubs.
  Thus, when the main algorithm stops, $r \le \opt$, and the output is a
  $(2+\epsilon')$-approximation, for a suitable $\epsilon'$.

  Assume $H^*$ is a solution that satisfies each demand with cost $r$ with
  minimum size.
  Recall $\troot$ is the root of the tree decomposition and $\cempty$ is the
  coloring of an empty bag.
  If $A_{\troot}(\cempty) \leq k$, then
  Lemma~\ref{lemma:algorithm_outputs_solution} states that the dynamic
  programming algorithm outputs a set of hubs $H$ of size at most $k$ that
  satisfies each demand in $D_{\troot}(\cempty) = \demands$ with cost
  $(1+\epsilon)2r$.
  Otherwise, $k < A_{\troot}(\cempty)$, and Lemma~\ref{lemma:not_many_hubs}
  implies $k < A_{\troot}(\cempty) \leq |H^* \cap V(G_{\troot})| = |H^*|$.
  Thus, by the minimality of $H^*$, there is no solution of cost~$r$ that opens
  $k$ hubs.

  Finally, we bound the running time.
  Let $n = |V(G)|$. The tree decomposition has $\OO(\tw \cdot n)$ nodes and, for
  each node $t$, the number of colorings is $|\Sigma|^{\OO({\tw})}$.
  Also, each recurrence can be computed in time  $\OA(|\Sigma|^{\OO(\tw)})$.
  Since $r = \OO(\frac{1}{\epsilon}n)$ and $\delta = \Theta\big(\frac{\epsilon}{\tw\cdot\log n}\big)$, the size of
  $\Sigma$ is
  \begin{align*}
    |\Sigma|
     & =
    \OO\left(\log_{1+\delta} r\right)
    =
    \OO\left(\frac{\log r}{\log(1+\delta)}\right)
    =
    \OO\left(\frac{\log n + \log (1/\epsilon)}{\delta}\right) \\
     & =
    \OO\left((\tw/\epsilon) \, (\log^2 n + \log n \log (1/\epsilon)) \right)
    =
    \OO\left((\tw/\epsilon)^2 \, \log^2 n  \right).
  \end{align*}
  Notice that $\OO(\log^{\OO(\tw)} n) = \OA(2^{\OO(\tw)})$, thus the total
  running time is bounded by $\OA\left(|\Sigma|^{\OO(\tw)}\right) =
  \OA\left((\tw/\epsilon)^{\OO(\tw)}\right)$.\qed
\end{proof}

%% file: content/remarks.tex
\section{Final remarks}
\label{section:remarks}

Our parameterized $(2+\epsilon)$-approximation algorithm circumvents hardness
barriers coming from both classical and parameterized complexity theories.
Improving on the $3$-approximation is \np-hard and, as we note, $\wtwo$-hard
even if we take $r$ as a constant and parameterize by $k$. Thus, since we drop
$k$ as parameter and take $r$ as part of the input, parameterizing by treewidth
is a necessary condition of the algorithm to break the $3$-approximation lower
bound. Approximating is also necessary, as the problem on planar graphs is
$\wone$-hard for pathwidth and several other parameters.

These results are analogous to \kcfull, which has a $2$-approximation lower
bound and does not admit an \fpt algorithm. Unlike \kcfull, however, we left
open whether \makhc admits an EPAS when parameterized by treewidth.
The challenge seems to be the non-locality of the paths serving the demands,
thus established techniques are not sufficient to tackle this issue.
In this paper, we show how to compute a special subset of demands that must be
served locally for each subproblem. We hope this technique may be of further
interest.
A possible direction of research is to consider the single allocation variant in
the two-stop model, which is a well-studied generalization of
\makhc~\cite{FarahaniHAN13,BeneditoP19}.

%% file: content/planar.tex
\section{The planar case}
\label{section:planar}

\newcommand{\vext}{V_{ext}}
\newcommand{\vint}{V_{int}}

In this section, we give a $(2+\epsilon)$-approximation algorithm parameterized
by $k$~and~$r$, when the input is restricted to unweighted planar graphs.
This algorithm can be seen as another way to challenge the approximation lower
bound presented in Section~\ref{section:hardness}.
Indeed, by Theorem~\ref{theorem:inapprox_in_fpt}, finding a
$(3-\epsilon)$-approximation parameterized by $k$ and $r$ is \wtwo-hard for
unweighted graphs, even when $r$ is a constant.
Thus, we restrict the input to planar graphs, but get a better approximation factor.

The algorithm is built upon the bidimensionality framework and follows the
arguments for \kcfull by Demaine~\etal~\cite{DemaineFHT05}.
In the following, let $(G,\clients,\hubs,\demands,k,r)$ be a positive instance
of \makhc such that $G$ is an unweighted planar graph.
\begin{lemma}
    \label{lemma:boundk}
    If $G$ has a $(\rho \times \rho)$-grid as minor, then $\rho \leq
    \sqrt{k}(2r+1) + 2r$.
\end{lemma}

\begin{proof}
    We begin with a series of definitions.

    Let $F$ be a $(\rho \times \rho)$-grid, where
    \begin{align*}
        V(F) & = \{1,\dots,\rho\} \times \{1,\dots,\rho\} \quad \mbox{and} \\
        E(F) & = \{((x,y),(x',y')) : |x-x'| + |y-y'| = 1\}.
    \end{align*}

    Let $\vext$ be the set of vertices of $F$ whose degrees are smaller
    than~$4$. We assume the vertices of $\vext$ belong to the external face of
    some embedding of~$F$ and call the other faces \emph{internal}.

    Let $\vint$ be the set of vertices of $F$ that have distance at least $r$
    from every vertex in $\vext$. Note that $\vint$ induces a subgraph
    $F[\vint]$ that is a subgrid of $F$ with $|\vint| = {(\rho - 2r)}^2$.

    Define $\delta((x,y), (x',y')) = \max\{|x-x'|, |y-y'|\}$ and let $d_F((x,y),
    (x', y'))$ be the distance from $(x,y)$ to $(x',y')$ in $F$.

    Let $J$ be the supergraph of $F$ with the same set of vertices and with the
    additional set of (diagonal) edges:
    \begin{align*}
        \{
        ((x, y), (x+1, y+1)),
        \,\,\,
        ((x, y+1), (x+1, y))
        \,:\,
        1 \leq x,y < \rho
        \}.
    \end{align*}

    Let $R$ be a subgraph of $J$ and let $d_R(u,v)$ be the length of a shortest
    path from $u$ to $v$ in $R$. Observe that, for every $u,v \in V(R)$, we have
    $\delta(u,v) \leq d_R(u,v)$. Define $N_R^\ell(u) = \{v : d_R(u,v) \leq \ell\}$.

    For a $(x,y) \in V(F)$ and an integer $\ell$, define
    \[
        B_\ell((x,y)) = \{(x',y') : \delta((x,y),(x',y')) \leq \ell\}.
    \]

    Now, consider a sequence of edge contractions and removals which transforms
    $G$ into a minor isomorphic to $F$ using a maximal number of edge
    contractions. Let $H$ be the result of applying only the contractions of
    that sequence to $G$, and consider an embedding of $H$ in the plane that
    corresponds to an embedding of~$F$.
    Partition the edges of $H$ in three sets: the edges that occur in $F$, the
    set $E_1$ that connect non-adjacent vertices of an internal face of $F$, and
    the set $E_2$ with all other edges. Note that edges in $E_2$ are only
    incident with vertices in~$\vext$.

    Call $R$ the graph we obtain by adding edges $E_1$ to $F$, and note that $R$
    is a subgraph of $J$. Then, for a vertex $u$ of $R$ and an integer $\ell$,
    we have that ${N_R^\ell(u) \subseteq B_\ell(u)}$.
    Observe that the set of edges of $H$ is $E(R) \cup E_2$. For a vertex $u \in
    \vint$, we claim that ${N_H^r(u) \subseteq B_r(u)}$. This holds because
    paths of length at most $r$ starting at a vertex of $\vint$ do not use edges
    of $E_2$ and, as a consequence, $N_H^r(u) = N_R^r(u)$.

    Let $S$ be a solution for the instance of \makhc. Observe that the distance
    between every client and a hub of $S$ is at most $r$, since every vertex is
    in some set $V(G_{ab})$. Also, note that, for vertices $u$ and $v$ of $G$
    associated with vertices $u'$ and $v'$ of $H$, $d_H(u',v') \leq d_G(u,v)$,
    as $H$ is obtained from $G$ using only edge contractions.

    Define a set of vertices:
    \[
        Y = \vint \cap \{((2r+1)i + r + 1),(2r+1)j + r + 1): i,j \in \integersnn\}.
    \]

    The size of this set is
    $
    |Y| \geq {\Big\lceil\frac{\rho - 2r}{2r+1} \Big\rceil}^2 \geq
    {\Big(\frac{\rho - 2r}{2r+1} \Big)}^2.
    $

    For distinct $y,y' \in Y$, we have $B_r(y) \cap B_r(y') = \emptyset$. Also,
    there must exist a hub in $S$ that is associated with some vertex in
    $N_H^r(y) \subseteq B_r(y)$. Therefore, each $y \in Y$ is associated to one
    unique hub in $S$, and finally,
    \begin{align*}
        k \geq |Y| \geq {\Big(\frac{\rho - 2r}{2r+1} \Big)}^2.\tag*{\qed}
    \end{align*}
\end{proof}

\begin{corollary}
    $\tw(G) \leq 6\sqrt{k}(2r+1) + 12r + 1$.
\end{corollary}

\begin{proof}
    Robertson, Seymour and Thomas\footnote{Theorem 6.2, N. Robertson, P. Seymour,
        and R. Thomas. Quickly excluding a planar graph. Journal of Combinatorial
        Theory, Series B, 1994.} prove that, if $G$ has no $((\rho+1) \times
    (\rho+1))$-grid as a minor, then $\tw(G) \leq 6(\rho + 1) - 5$.
    Let $\rho$ be the largest integer for which $G$ has a $(\rho \times
    \rho)$-grid as a minor. Then, using Lemma~\ref{lemma:boundk}, we have that
    $\tw(G) \leq 6\sqrt{k}(2r+1) + 12r + 1$.\qed
\end{proof}

Using the previous bound and Theorem~\ref{theorem:main}, we get the main result
of this section.

\begin{theorem}
    \label{theorem:planar}
    For every $\epsilon > 0$, there is a parameterized
    $(2+\epsilon)$-approximation algorithm for \makhc when the parameters are
    $k$ and $r$, and the input graph is unweighted and planar.
\end{theorem}

Notice that a version of the dynamic programming algorithm presented in
Section~\ref{section:algorithm} that stores distances exactly is a
$2$-approximation parameterized by $\tw$ and $r$. Thus, for \makhc with
unweighted planar graphs, there is actually a $2$-approximation algorithm
parameterized by $k$ and $r$.